\documentclass[twocolumn,preprintnumbers,amsmath,amssymb,floatfix,letter]{revtex4-1}

\usepackage[margin=1.0in]{geometry}
\usepackage{wrapfig}
\usepackage{times}
\usepackage{epsfig}
\usepackage{graphicx}
\usepackage{color}
\usepackage{dcolumn}
\usepackage{bm}
\usepackage{amsmath}
\usepackage{setspace}
\usepackage[center]{subfigure}

\newcommand{\beq}{\begin{equation}}
\newcommand{\eeq}{\end{equation}}

\newcommand{\beqa}{\begin{eqnarray}}
\newcommand{\eeqa}{\end{eqnarray}}

\begin{document}

\title{Characterizing Solute Segregation and Grain Boundary Energy in Binary Alloy Phase Field Crystal Models}

\author{Jonathan Stolle}  
\email{stolle@mcmaster.ca} \affiliation{Department of Physics and Astronomy, McMaster University, 1280 Main Street West,
Hamilton, Ontario L8S-4L7}

\author{Nikolas Provatas}
\email{provatas@physics.mcgill.ca} \affiliation{Department of Physics, Centre for the Physics of Materials, McGill University, 3600 rue University,
Montreal, Quebec H3A-2T8} \affiliation{Department of Materials Science and Engineering, McMaster University, 1280 Main Street West,
Hamilton, Ontario L8S-4L7}

\date{today}

\begin{abstract}
This paper studies how solute segregation and its relationship to grain boundary energy in binary alloys is captured in the phase field crystal (PFC) formalism, a continuum method that incorporates atomic scale elasto-plastic effects on diffusional time scales.  Grain boundaries are simulated using two binary alloy PFC models --- the original binary model by Elder et al (2007) and the XPFC model by Greenwood et al (2011).   In both cases, grain boundary energy versus misorientation data is shown to be well described by Read-Shockley theory.  The Gibbs Adsorption Theorem is then used to derive a semi-analytic function describing solute segregation to grain boundaries. This is used to characterize grain boundary energy versus average alloy concentration and undercooling below the solidus.  We also investigate how size mismatch between different species and their interaction strength affects segregation to the grain boundary.  Finally, we interpret the implications of our simulations on material properties related to interface segregation. 
\end{abstract}

\maketitle
\section{Introduction}
\label{secIntro}
Microstructure in metals is important for determining many of  their properties (e.g., mechanical, thermal, electrical).  The various defects associated with microstructure formation (e.g., grain boundaries, dislocations, vacancies) contribute to an excess of free energy of a system.  As a material evolves towards equilibrium, its microstructure changes and along with it the material's properties. Grain boundaries are among the most important defects in metals. Their energy, composition, and distribution directly affect the flow of dislocations and influence the thermodynamics of second phase and precipitate formation. 

In alloys, segregation of solute atoms can alter grain boundary energy  \cite{cahnphysical1983, sutton_interfaces_1995, lejekthermodynamics1995}. The effect of segregation can also manifest itself in other ways.  Two other properties strongly affected by solute segregation are solute drag \cite{hoyt_2010, PhysRevB.76.024111, sutton_interfaces_1995} 
and grain boundary wetting \cite{Mishin20093771, PhysRevE.81.051601, luo_stabilization_2007, sutton_interfaces_1995}.  In the former case, the grain boundary energy is reduced by solute segregation, thus reducing the driving force to reduce surface area (excess free energy) of a grain boundary. In the latter case, solute segregation can dramatically affect the thermodynamics of grain boundary formation; not only can segregation alter at what undercooling grain boundary wetting occurs, but it can allow for different grain boundary states (e.g., grain boundaries widths) \cite{PhysRevE.81.051601}.

There have been a number of experimental studies of grain boundary energy involving pure materials 
\cite{Gjostein1959319,austandchalmers, PhysRev.78.275} and alloys \cite{PhysRev.78.275, cahnphysical1983}.  Many 
studies have focused on characterizion of solute segregation and distribution 
\cite{sutton_interfaces_1995, lejekthermodynamics1995} as solute segregation typically has an important effect on 
grain boundary energy as demonstrated in \cite{cahnphysical1983, Kirchner2011406, Shibuta20091025}. For pure materials and dilute alloys, many 
of these studies have found that the grain boundary energy is well-fit by the well-known Read-Shockley Law when 
neighbouring grains are misoriented by small angles \cite{PhysRev.78.275,Gjostein1959319}. It is also possible to 
adjust the parameters of the Read-Shockley equation to fit a larger range of misorientation angles 
\cite{PhysRev.78.275, Gjostein1959319, austandchalmers}.  

A number of theoretical approaches have also studied grain boundary energy in metals.  The most prevalent, for both pure materials and alloyed metals, are the analytic and semi-analytic dislocation models of Read and Shockley \cite{PhysRev.78.275, hirth_theory_1992} and Van der Merwe \cite{sutton_interfaces_1995}, and models employing simple thermodynamic considerations of an interface \cite{sutton_interfaces_1995, Kirchner2011406}.  Various computational approaches have also been employed to determine grain boundary energy in pure metals, including Monte Carlo simulations \cite{PhysRevB.49.14930} and lattice statics \cite{PhysRevB.54.R11133}.  Some computational approaches have also been used to model solutal effects in grain boundaries. These include monte carlo methods \cite{lejekthermodynamics1995}, molecular dynamics \cite{Shibuta20091025,lejekthermodynamics1995}  and phase field simulations \cite{PhysRevE.81.051601}.

A relatively new continuum approach for modelling  the effect of defects in non-equilibrium phase transitions has emerged in the past ten years known as the phase field crystal (PFC) method. The PFC methods have been developed as part of a continued attempt to bridge the divide between traditional phase field modelling and molecular dynamics approaches \cite{PhysRevE.81.011602}. In particular, PFC methods access diffusional time scales while incorporating the salient features of atomic-scale elasticity, plasticity \cite{PhysRevB.75.064107,Jin06,Fallah12} and dislocation properties \cite{Berrydefect2}.  In pure materials, PFC simulations of grain boundary energy have yielded excellent correspondence with the Read-Shockley theory of  grain boundary energy versus misorientation \cite{eldermodeling2004, mellenthinphasefield2008}. They have also shed light on the physics of grain boundary pre-melting in pure materials \cite{Berry_2008,mellenthinphasefield2008}. To date, the grain foundry energetics alloys simulated by PFC models has not been characterized. 

This work systematically characterizes the thermodynamics of grain boundary segregation and grain boundary energy in two binary alloy PFC models, the first the original PFC  model of Ref.~\cite{PhysRevB.75.064107} and the structural PFC model of Ref.~\cite{Greenwood2011}. Section (\ref{secPFCModel}) introduces the two PFC models used in this study  and reviews the Gibbs Adsorption theorem (\ref{secThermoGB}).  Section \ref{secNumericalResults} reports on numerical simulations of the aforementoned PFC models that characterize grain boundary energy and compares computed solute adsorption to the prediction of the Gibbs adsorption theorem. Our results are discussed in the context of previous experiments and theories. Section \ref{secDiscussion} discusses the effect of different model parameters on our results.  Section~\ref{secConclusion} concludes and summarizes our study.

\section{Phase Field Crystal Model of a Binary Alloy} 
\label{secPFCModel}
\subsection{Original Binary PFC Model} 
\label{secPFCModelOrig}
The original phase field crystal model (PFC) of alloys characterized in this work is derived in detail in Ref. \cite{PhysRevB.75.064107}. The resultant PFC free energy is expressed in terms of a temporally coarse-grained normalized crystal density field and a relative density difference that is analogous to a solute concentration field. In particular, the normalized total density is given by $n=(\rho-\rho_l)/\rho_l$ and the normalized concentration by $\psi=(\rho_1-\rho_2)/\rho_l$, where the total density $\rho$ is the sum of the density of each species, $\rho=\rho_1+\rho_2$, and $\rho_l$ is the density of a reference liquid state.  The dimensionless Helmholtz free energy functional expressed in these variables is given by 
\beqa
	F&=& \int_V  \left\{ (B^L_0+B^L_2\psi^2)\frac{n^2}{2}+B^X n (2 \nabla^2 + \nabla^4) \frac{n}{2}  \right. \nonumber \\
	&-&t\frac{n^3}{3}+ \left. v\frac{n^4}{4}+w\frac{\psi^2}{2}+u\frac{\psi^4}{4}+K\frac{|\nabla\psi|^2}{2} \right. \nonumber \\ &+& \left. \eta B^X n \psi (\nabla^2+\nabla^4 ) n \, d \vec{r} \right\}
	\label{eqPFC}
\eeqa
where $B^L_0$ is the isothermal compressibility of the liquid at $\psi=0$, $B^L_2$ determines how the isothermal compressibility of the liquid changes with $\psi$, $B^X$ is related to elastic constants in the solid, and $t$, $v$, $u$ are determined by the low order terms of a local expansion of the classical density functional theory description of the material, $w$ is related to the atomic bond energies, and $K$ is related to $w$ and the lattice spacing \cite{PhysRevB.75.064107}.  The difference $B^L_0-B^X$ plays the role of normalized temperature variable. All lengths are scaled such that the lattice constant is $a=4\pi/\sqrt{3}$ when the lattice mismatch parameter $\eta=0$. The lattice spacing changes with concentration according to the parameter $\eta=\left(1/a\right) \partial a /\partial \psi$.

Assuming conserved dissipative dynamics for both fields, the evolution equations are:
\beqa
	{\frac{\partial n}{\partial t}=\nabla^2\left(\frac{\delta F}{\delta n}\right)=\nabla^2\mu_n
	\label{eqnevol}	}
	\\ {\frac{\partial \psi}{\partial t}=\nabla^2\left(\frac{\delta F}{\delta \psi}\right)=\nabla^2\mu_\psi
	\label{eqpsievol}	}
\eeqa
In Eqs.~\ref{eqnevol} and ~\ref{eqpsievol}, the constant atomic mobilities have been absorbed in the time variable and a noise term reflecting the effect of thermal fluctuations on the evolution of the system has been neglected.  The chemical potentials 
corresponding to each conserved field are given by $\mu_n$=$\delta F/{\delta n}$ and $\mu_{\psi}$=$\delta F/{\delta \psi}$.

Because we are solely analyzing thermodynamic aspects of the system, the equilibrium states can be found more quickly and accurately by solving:
\beqa
	{\frac{\partial n}{\partial t}=-\left(\frac{\delta F}{\delta n}-\mu_n\right)
	\label{eqnevolA}	}
	\\ {\frac{\partial \psi}{\partial t}=-\left(\frac{\delta F}{\delta \psi}-\mu_\psi\right)
	\label{eqpsievolA}	}
\eeqa

where the chemical potentials, $\mu_\psi$ and $\mu_n$, are thermodynamic control parameters, analogous to temperature and pressure, and $t$ is pseudotime; this formalism is used for a pure material in \cite{mellenthinphasefield2008}.

Equations~\ref{eqPFC}-\ref{eqpsievol} can be represented on mesoscales by a set of complex order parameter 
equations, the coefficients of which are directly linked to those of the above PFC model, which is,  in turn, linked 
to a simplified classical density functional theory of freezing. The complex order parameter representation 
of Eqs.~\ref{eqPFC}-\ref{eqpsievol}  has also been shown to reduce to the form of a traditional scalar phase field model with coupled strain effects \cite{PhysRevE.81.011602}. To the accuracy of a single-mode approximation,  such an analysis 
thus yields a microscopic connection between continuum elastic effects and solute concentration and temperature. 

\subsection{Binary XPFC Model} 
\label{secPFCModelXPFC}
The second phase field crystal (PFC) model of binary alloys characterized in this work is derived in detail in Ref. \cite{Greenwood2011}. The resultant PFC free energy is expressed in terms of a temporally coarse-grained normalized crystal density field and a solute concentration field. In particular, the normalized total density is given by $n=(\rho-\rho_0)/\rho_0$ and the concentration field by $c=\rho_2/(\rho_1+\rho_2)$, where the total density $\rho$ is the sum of the density of each species, $\rho=\rho_1+\rho_2$, and $\rho_0$ is the density of a reference state.  The dimensionless Helmholtz free energy functional expressed in these variables is given by 

\beqa
	F&=& \int_V  \left\{ \frac{n^2}{2}  -\eta \frac{n^3}{6}+  \chi \frac{n^4}{12}   \right. \nonumber \\
	&+&\left. \omega (c \ln(c/c_0)+ (1-c) \ln((1-c)/(1-c_0))) \right. \nonumber \\ 
	&+& \left. \frac{n}{2} \int_{V'}{\left(C_{eff}(\vec{r}-\vec{r}') n(\vec{r}') d\vec{r}'\right)}  + \alpha \frac{|\nabla c|^2}{2}\, d \vec{r} \right\}
	\label{eqXPFC}
\eeqa
where $\eta$ and $\chi$ are prefactors chosen to fit the ideal free energy,  $\omega$ and $c_0$ determines the strength of the entropy of mixing, and $\alpha$ is a constant related to the two-point correlation functions but taken here as a constant \cite{Greenwood2011}.  The correlation function $C_{eff}$ is given by  
\beqa
C_{eff}&=&g(c) C_{11} + (1-g(c)) C_{22},  \nonumber \\
 g(c) &=& 1- \lambda c + (3+\lambda)c^2+4 \nonumber
\eeqa
with $\lambda$ being the enthalpy of mixing in the solid state and $C_{xy}$ is the correlation function between species $x$ and $y$.   The fourier transform of this correlation function is given by:

\beqa
	\hat{C}_{xx}(k)=\sum^N_{i=1}{P_i\exp\left(-D_i\sigma^2 k_i^2\right) \exp\left(-G_i (k-k_i)^2\right)} \nonumber \label{eqinterp}
\eeqa
where $N$ is the total number of family of peaks, $\sigma$ is a variable representing the temperature, $k_i$ is the magnitude of the wave number of the family of peaks, $D_i$, $P_i$, and $G_i$ are free parameters for the $i^{th}$ family of planes, treated here for simplicity as adjustable constants.  Note that the lattice spacing is $a=2\pi/k_i$.

Assuming conserved dissipative dynamics for both fields, the evolution equations are:
\beqa
	{\frac{\partial n}{\partial t}=\nabla^2\left(\frac{\delta F}{\delta n}\right)=\nabla^2\mu_n
	\label{eqnevolX}	}
	\\ {\frac{\partial c}{\partial t}=\nabla^2\left(\frac{\delta F}{\delta c}\right)=\nabla^2\mu_c
	\label{eqcevolX}	}
\eeqa
Once again in Eqs. \ref{eqnevolX} and ~\ref{eqcevolX}, the atomic mobilities have been absorbed in the time variable and thermal fluctuations have been neglected.  The chemical potentials corresponding to each field are given by $\mu_n$=$\delta F/{\delta n}$ and $\mu_{c}$=$\delta F/{\delta c}$.

As in the original PFC model, the equilibrium states of the PFC model are found  found by solving
\beqa
	{\frac{\partial n}{\partial t}=-\left(\frac{\delta F}{\delta n}-\mu_n\right)
	\label{eqnevolXA}	}
	\\ {\frac{\partial c}{\partial t}=-M_c\left(\frac{\delta F}{\delta c}-\mu_c\right)
	\label{eqcevolXA}	}
\eeqa

where $\mu_c$ and $\mu_n$ are thermodynamic control parameters and $M_c$ is a relative relaxation mobility of the $c$ field against that of the $n$ field.

\section{Thermodynamics of Grain Boundaries}
\label{secThermoGB}
To investigate the thermodynamics of segregation behaviour at a grain boundary in a binary alloy, we need to consider grain boundary energy.  Grain boundary energy in pure materials can be determined by a number of different methods. For grain boundaries with a small misorientation angle, Read and Shockley derived the relation now named after them,
\beq
	\gamma_{gb}=E_0\theta(A-\ln(\theta)),
	\label{eqReadShockley}
\eeq
by considering the grain boundary as an array of dislocations, where the dislocation cores do not overlap.  For a 2D crystal, the constants in Eq.~\ref{eqReadShockley} are $E_0=Y_2 b/(8 \pi \alpha)$, where $Y_2$ is the 2D elastic modulus, $\alpha=\sqrt{3}/2$ is a correction factor for hexagonal as opposed to square geometry, $b \approx a$ is the Burger's vector of the dislocation, and $A=1+\ln(a/r_0)-\ln(2\pi)\approx1.5-\ln(2\pi)$, which is related to the core energy by the core radius, $r_0$ 
\cite{eldermodeling2004, mellenthinphasefield2008}.  For high angles, some theoretical approaches consider the forces between 
atoms in a fixed geometry \cite{PhysRevB.54.R11133, sutton_interfaces_1995}, while others treat the high angle grain boundary as an amorphous phase sandwiched between 2 bulk phases \cite{cahnphysical1983}.  However, as already noted in 
\cite{Gjostein1959319, austandchalmers, cahnphysical1983}, the parameters in Eq.~\ref{eqReadShockley} can be chosen to give 
a reasonable fit between for the relation between grain boundary energy and misorientation for much larger angles than those considered in the original problem. 

It is reasonable to assume that the form of Eq.~\ref{eqReadShockley} will remain valid for binary alloys, with the coefficients $E_0$ and 
$A$ modified by the presence of segregated solute, as well as by the degree of undercooling. This hypothesis is consistent with phase 
field and phase field crystal simulations of a pure material Ref.~\cite{war03,mellenthinphasefield2008}, which used Eq.~\ref
{eqReadShockley} to model grain boundary energy in pure materials at different undercoolings by fitting the core energy (i.e. parameter $A$) to temperature.

In comparing the theoretical forms of grain boundary energy and segregation to corresponding values computed directly form model simulations, we will be guided by the Gibbs' adsorption theorem, which relates the degree of solute segregation to the grain boundary energy and chemical potential of the system according to 
\begin{equation}
	\left(\frac{\partial \gamma_{gb}}{\partial \mu_x}\right)_{T,p}=-\Gamma^{ex}_{x}
	\label{eqGibbsads}
\end{equation} 
where $\mu_x$ is the chemical potential of species $x$ in the binary alloy ($x=1, 2$) and $\Gamma^{ex}_{x}=N^{ex}_x/A$, where 
$N^{ex}_x$ is the excess number of atoms of species $x$ in a unit area $A$ of the grain boundary (unit length in 2D) 
\cite{sutton_interfaces_1995}.  It is noted that Eq.~\ref{eqGibbsads} can be also be written in terms of the chemical potential difference, $\mu_{\psi}=\mu_1-\mu_2$, if $\Gamma_x^{ex}$ is replaced by $\Gamma_{\psi}^{ex}$, the excess particle difference at the grain boundary, as in the model by Elder et al \cite{PhysRevB.75.064107}.  Similarly when working with the XPFC model \cite{Greenwood2011}, Eq.~(\ref{eqGibbsads}) can be written by substituting in $\mu_c$ for $\mu_x$ and the excess concentration $\Gamma_{c}^{ex}$ for $\Gamma_{x}^{ex}$ .

Alternatively, because the chemical potential of the normalized number density is more easily controlled than pressure, the form of Gibbs' absorption theorem is often used in this report is:
\begin{equation}
	\left(\frac{\partial \gamma_{gb}}{\partial \mu_x}\right)_{T,\mu_n}=-\Gamma^{ex}_{x}
	\label{eqGibbsads2}
\end{equation}

\section{Numerical Results}
\label{secNumericalResults}

\subsection{PFC Model}
\label{secNumericalPFCOrig}
\subsubsection{Method}
\label{secPFCOrigMethod}

We simulate Eqs.~\ref{eqnevol} and \ref{eqpsievol} with the Fourier methods outlined in Elder and Grant \cite{eldermodeling2004} and Mellenthin et al \cite{mellenthinphasefield2008}. Our computations of Eqs~\ref{eqnevol} and \ref{eqpsievol} are performed on a 1024x2048 grid with periodic boundary conditions.  The grid spacing is $\Delta x = \pi/4 \sqrt{(1-2\eta\psi_s)/(1-4\eta\psi_s)}$ and the time step is $\Delta t = 1.0$.  In this work, the average normalized alloy concentrations studied are $\psi_0$=0.1,-0.05, -0.15, and -0.2 (-0.2 is only considered for the large angle fits), which differs a little from $\psi_s$ because of solute segregation to the grain boundary.  The average normalized density in the PFC model is set to $n_0=0$.  Bicrystal grain boundaries are studied in alloys whose equilibrium phase diagram is described by a spinodal phase diagram, i.e. $w=0.088$, except for alloys with $\psi_0=-0.2$, for which $w=0.008$ \cite{Provatas_Elder_book_2010}.  For each concentration studied, the parameter $B^L_0$ is chosen such that the undercooling is sufficient for grain boundaries to close (that is, the disjoining pressure does not keep the grain boundaries from closing as it could at too small undercooling 
\cite{mellenthinphasefield2008, PhysRevE.81.051601}).  The exact value for $B^L_0$ depends on the concentration, however, 
the undercooling typically varies from $-0.02$ to $-0.06$.  The other parameters used in Eqs.~\ref{eqnevol} and ~\ref{eqpsievol} are $B^L_2=-1.8$, $B^X=1$, $t=0.6$, $v=1$, $u=4$, and $K=4$.  Note that $\eta=0$ except where otherwise indicated.

\begin{figure}
\includegraphics[width=3.5cm]{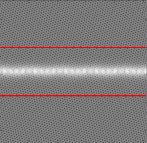}
\includegraphics[width=3.5cm]{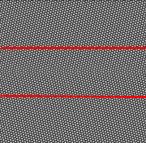}
\includegraphics[width=3.5cm]{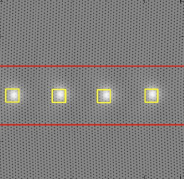}
\includegraphics[width=3.5cm]{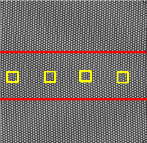}
	\caption{Images of concentration \emph{(left frames)} and density fields \emph{(right frames)} around low angle \emph{(bottom frames)} and high angle \emph{(top frames)} grain boundaries.  The control area enclosing the grain boundary lies between the two red lines.  The dislocations in the low angle grain boundary are shown with yellow squares.  Note that the x-direction in simulations is the vertical direction in the figures.}
	\label{fig:gb_image}
\end{figure}

The basic initial condition for simulating a grain boundary  begins with two large grains with a small liquid gap of roughly 10 grid points in between them.  For the least deep temperature ($B_0^L$) quenches, at a given average concentration $\psi_0$, the temperature parameters is dropped and a grain boundary forms. Quenches to the lowest values of $B_0^L$ are done in multiple steps, quenching first to higher, intermediate temperature, before lowering $B_0^L$ to the final desired value.  The crystals are oriented $\theta/2$ and $-\theta/2$ from the 0$^\circ$ crystal reference, respectively.  Once a simulation is started, the grains quickly form a bicrystal with misorientation, $\theta$.   The grain boundary normal is in the $x$-direction.  The angles chosen were: 1.55$^\circ$, 2.07$^\circ$, 2.58$^\circ$, 3.10$^\circ$, 4.14$^\circ$, 5.17$^\circ$, 6.20$^\circ$, 8.23$^\circ$, 10.3$^\circ$, 12.5$^\circ$, 15$^\circ$, 17.5$^\circ$, 20$^\circ$, 22.5$^\circ$, 25$^\circ$, 27.5$^\circ$, 30$^\circ$, 32.5$^\circ$, 35$^\circ$, 40$^\circ$.  The small angles are chosen such that an integral number of evenly-spaced dislocations fit within the numerical domain, so that the results can be compared accurately to Eq.~\ref{eqReadShockley}.  Due to periodic boundary conditions, two grain boundaries form.  The initial spacing between the two grain boundaries is 512 grid points, which is chosen such that interaction between two grain boundaries is negligible (except at possibly 1.55$^\circ$).  Examples of a low angle and a high angle grain boundaries are shown in Fig. \ref{fig:gb_image}. 

Grain boundary simulations at each $B_0^L$ were typically run for 100000 time steps.  Equilibration of a grain boundary is determined by the standard deviation of the chemical potential, $s_{\mu_x}$, and its magnitude relative to the mean of the chemical potential, $\mu_x$, where $x$ is $\psi$ or $n$.  If either $s_{\mu_c}$ or $s_{\mu_n}$ is much greater than $10^{-5}$ or $s_{\mu_c}/\mu_c$ or $s_{\mu_n}/\mu_n$ is greater than $10^{-2}$, simulations 
are run longer, until the above criteria are met. 

\subsubsection{Calculation of Grain Boundary Energy}
\label{secPFCOrigGBenergy}
Grain boundary energy is given by the excess grand potential per unit area (or length in 2D),
\beqa
	L_y \gamma_{gb}&=&L_y w_{gb}\left( f_{gb}-\mu_{\psi} \psi_{gb}-\mu_{n} n_{gb} \right. \nonumber \\
	&-& \left. \left(f_{s}-\mu_{\psi} \psi_s-\mu_{n} n_s \right)\right)
	\label{eqgrandpot}
\eeqa
where $L_y$ is the length of the grain boundary, $w_{gb}$ is the width of a region encompassing the grain boundary, $f_{gb}$ is the free energy density of the a domain enclosing the grain boundary,  $f_{s}$ is the free energy density of the bulk solid, $\psi_{gb}$ and $\psi_{s}$ are the average normalized concentrations in the grain boundary region and in the bulk solid, respectively, while $n_{gb}$ and $n_{s}$ are the average normalized atomic densities in the grain boundary region and in the solid, respectively.  $\psi_{gb}$ and $\psi_s$ and the corresponding normalized density quantities (obtained by substituting $\psi$ with $n$) are calculated as follows:
\beqa
\psi_{gb}&=&\frac{1} {L_y {w_{gb}}}  \int_{0}^{L_y} {\int_{x_{gb}-w_{gb}/2}^{x_{gb}+w_{gb}/2} {\psi dx} dy}
\nonumber \\
\psi_{s}&=&\frac{0.5 L_x \psi_0 - w_{gb}\psi_{gb}}{0.5 L_x-w_{gb}}
\nonumber
\eeqa
Since $\psi_s$ and $n_s$ do not differ substantially from the average values $\psi_0$ and $n_0$, respectively, $f_s$ can be written as a Taylor series about $n_0$ and $\psi_0$ as per \cite{mellenthinphasefield2008}, which gives
\beqa
	\gamma_{gb}&=&w_{gb}\left\{ f_{gb}-\mu_{\psi} (\psi_{gb}-\psi_{0}) \right. \nonumber \\
	&-& \left. \mu_{n} (n_{gb}-n_0) - f_s(n_0,\psi_0) \right\}
	\label{eqgrandpot0}
\eeqa
where $f_s(n_0,\psi_0)$ is explicitly written in terms of the bulk solid at density $n_0$ and concentration $\psi_0$, and  
\beq
f_{gb}=\frac{1} {L_y w_{gb}}  \int_{0}^{L_y} {\int_{x_{gb}-w_{gb}/2}^{x_{gb}+{w_{gb}}/2} {f dx} dy} \nonumber
\eeq

To identify  $w_{gb}$, the region encompassing the grain boundary, the position along the x-axis (transverse to the grain boundary) with the maximum average free energy density is first found ($x_{gb}$).  To determine the excess concentration and density, we determine the average value of the field summed over all positions within $\pm$ 80 grid points of this reference position (n.b., the effective width of the grain boundary is thus taken as $w_{gb} = 161 \Delta x$), because only small differences were observed when calculating the excess quantities with larger $w_{gb}$.
Following \cite{mellenthinphasefield2008}, we take $w_{gb}=L_x$, meaning that both boundaries are encompassed in the calculation Eq~\ref{eqgrandpot0}, which gives to first order:
\beqa
	2\gamma_{gb}&=&L_x \left\{ f_{gb}-\ f_s(n_0,\psi_0) \right\}
	\label{eqgrandpot1}
\eeqa

The calculation of grain boundary energy is made difficult by the accurate determination of $f_s$, as noted by \cite{mellenthinphasefield2008}. To determine this quantify we proceeded as follows. 
For each quench temperature $B^L_0$ and average alloy composition $\psi_0$, Eq.~\ref{eqReadShockley} was substituted for $\gamma_{gb}$ in Eq.~\ref{eqgrandpot1}. The parameters $f_s$ and $A$ were found by fitting the resulting expression for 
grain boundary energy versus misorientation to our computed data. Two types of fitting were performed.  The first case considered only low angles. We assumed a theoretical value for the elastic modulus $E_0$ analogously to Ref.~\cite{mellenthinphasefield2008} and 
obtained a best fit for $f_s$ and $A$ (or alternatively $r_0$).  The elastic modulus in this case was taken from the analytic expression 
derived in Refs.~\cite{eldermodeling2004, PhysRevE.81.011602} using a one-mode approximation to the total 
density, i.e. by writing  $n=\phi \sum_j \exp\left( \vec{G}_j \cdot \vec{x}\right)$, where $\vec{G}_j$ are the reciprocal lattice vectors of the crystal symmetry being considered (here 2D HCP).  As shown in Ref.~\cite{PhysRevE.81.011602}, $Y_2$ for a binary alloy is given by
\beqa
   Y_2 \! &=&\! 8 B^X \phi^2
   \label{eqPFCelast}
   \\ \nonumber
   \phi \! &=&\! \frac{t \!+\! \sqrt{t^2 \!-\! 15v \, \left\{\Delta B_o \!+\! (B^L_2-4 B^X_0 \eta^2)\psi_0^2\right\} }}{15 v}
\eeqa
where $\Delta B_o \! \equiv \! B^L_0 - B^X$ and where $\phi$ is the equilibrium amplitude of the first modes of the total density $n$.   

\begin{figure}
	\includegraphics[scale=0.45]{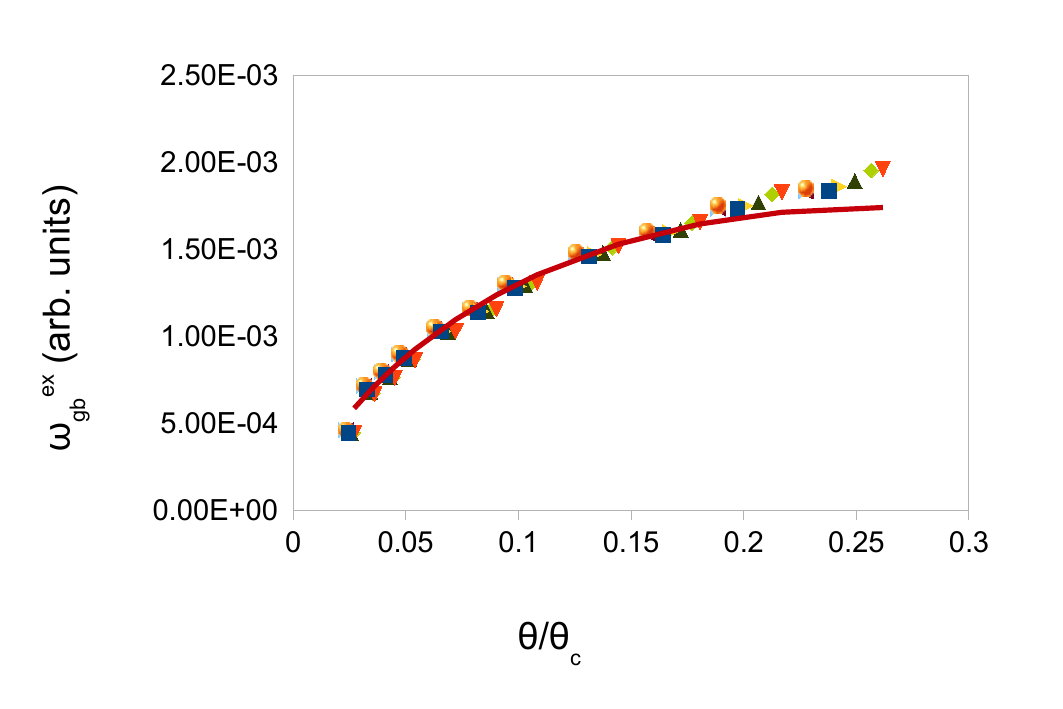}
	\caption{Scaled grain boundary energy for the alloy PFC model. Plotted are $\omega_{gb} = \gamma_{gb} \Delta x \theta_c /(w_{gb} \phi^2)$, vs. normalized bicrystal crystal misorientation (in radians for the reference curve), $\theta/\theta_c$, for different concentrations and temperatures for low angles, where $\theta_c=r_0\exp(0.5)/a$ is a horizontal stretch factor.   Reference Read-Shockley curve (solid); $\psi_0=0$, $B^L_0=1.002$ (squares); $\psi_0=0$, $B^L_0=0.962$ (inverted triangle); $\psi_0=-0.05$, $B^L_0=0.996$ (right triangle); $\psi_0=-0.05$, $B^L_0=1.006$ (circle); $\psi_0=0.1$, $B^L_0=1.015$ (left triangle); $\psi_0=0.1$, $B^L_0=0.995$ (star); $\psi_0=-0.15$, $B^L_0=1.035$ (bow tie); $\psi_0=-0.15$, $B^L_0=1.015$ (triangle); $\psi_0=-0.15$, $B^L_0=0.005$ (diamond).}
	\label{fig:gbenergysmall}
\end{figure}

\begin{figure}
	\includegraphics[scale=0.45]{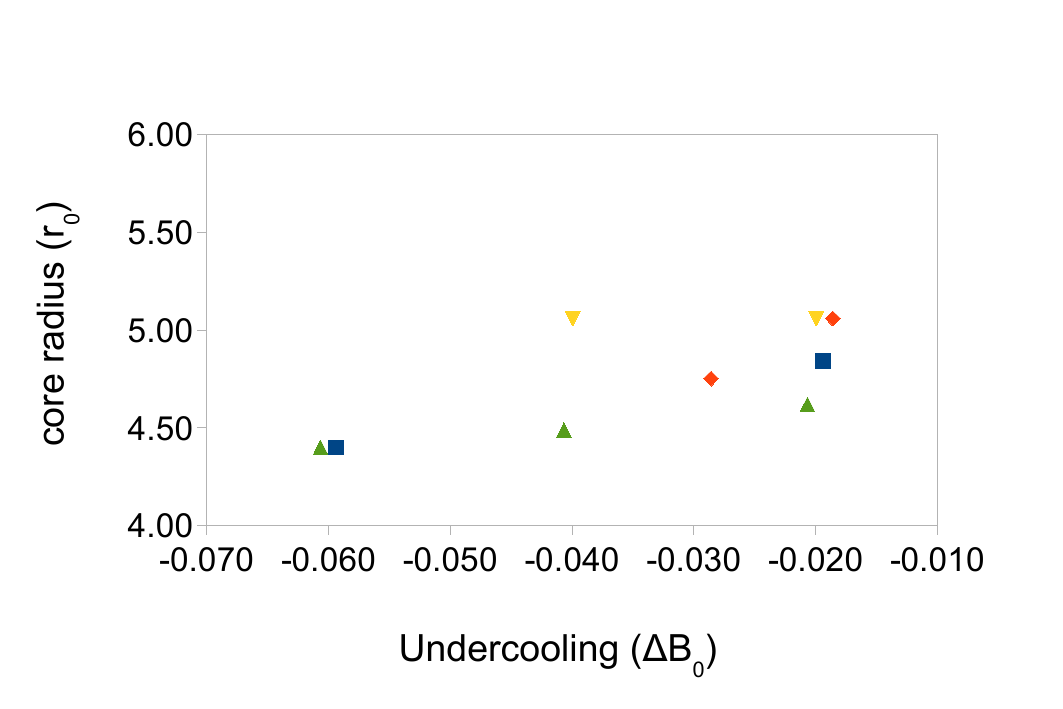}
	\caption{Core radius versus undercooling for different concentrations calculated from low angle fit of data in Fig. \ref{fig:gbenergysmall}.  Square (blue) - $\psi_0=0$; Diamond (red) - $\psi_0=-0.05$; Upside down triangle (yellow) - $\psi_0=0.1$; Triangle (green) - $\psi_0=-0.15$}
	\label{fig:gbcoreradius}
\end{figure}

Figure~\ref{fig:gbenergysmall} compares Equation~\ref{eqReadShockley} to grain boundary energies computed directly by Eq.~(\ref{eqgrandpot1}), for bi-crystals simulated with $w=0.088$.  The coefficients $E_0$, $A$ and $f_s$ were determined as described above.  To ensure that all points are fit to the normalized version of Eq. \ref{eqReadShockley}, the datasets are all compressed/stretched by a factor $\theta_c=r_0\exp(0.5)/a$, while $E_0=Y_2 b /(8 \pi \alpha \phi^2 w_{gb})$ and $A = 1.5-ln(2\pi)$, the latter of which is equivalent to $r_0 \approx 4.4$.  The core radii, plotted in Fig. \ref{fig:gbcoreradius}, can be seen to be relatively constant, though possibly decreasing for larger undercoolings $\Delta B \equiv B^L_0-B^L_{0s}$ ($B^{L}_{0s}-B^{X}_{0}$ defines the temperature of the solidus at a given concentration), consistent with the trend found in  \cite{mellenthinphasefield2008} for pure materials.   Note that the results of this method were verified against the method in \cite{mellenthinphasefield2008} for determining grain boundary energy and were found to be nearly identical.  


\begin{figure}
	\includegraphics[scale=0.4]{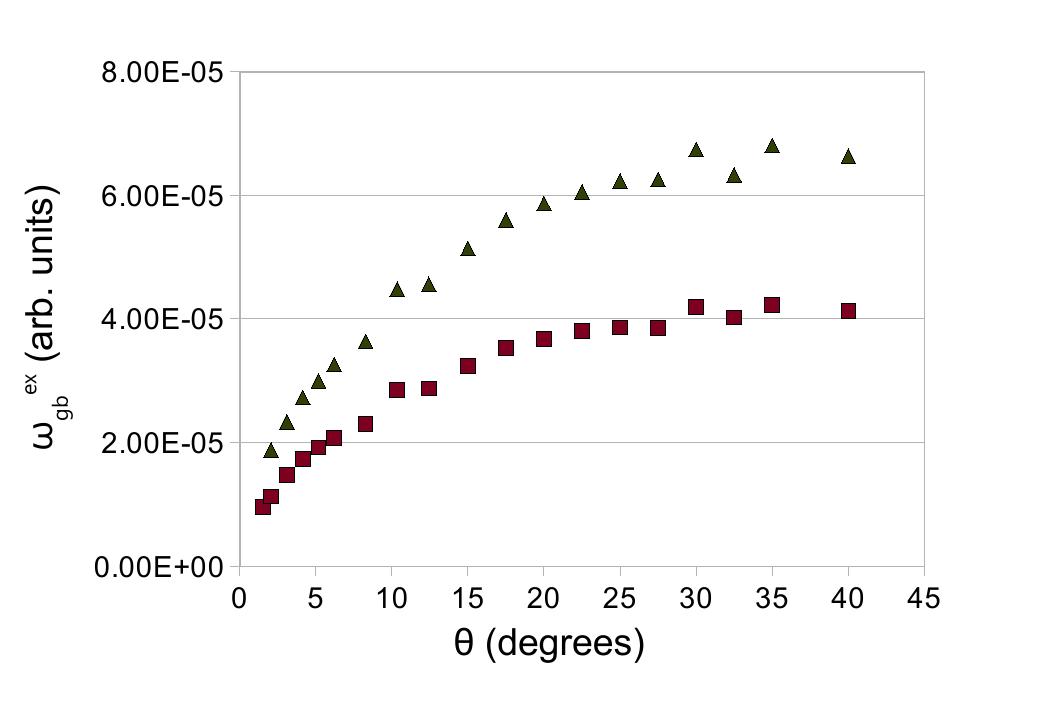}
	\caption{Normalized grain boundary energy for the PFC model for different undercooking. Plotted are $\omega_{gb}=\gamma_{gb} \Delta x /w_{gb}$) vs. misorientation at $B^L_0=1.015$ for $\psi_0=0.1$ (squares) and $\psi_0=-0.15$ (triangles). 
	}
	\label{fig:gb_energy_isothermal}
\end{figure}

Using the $f_s$ calculated from the the low-angle Read-Shockley data in Fig.~\ref{fig:gbenergysmall}, we also can explicitly examine how undercooling affects the grain boundary energy.  We examine the effect of undercooling for different concentrations at a given temperature. Simulations with larger $|\psi_0|$ represent larger undercoolings as given by the phase diagram in \cite{PhysRevB.75.064107}.  Undercooling has a very strong effect as illustrated in Fig.~\ref{fig:gb_energy_isothermal}.   

\begin{figure}
	\includegraphics[scale=0.45]{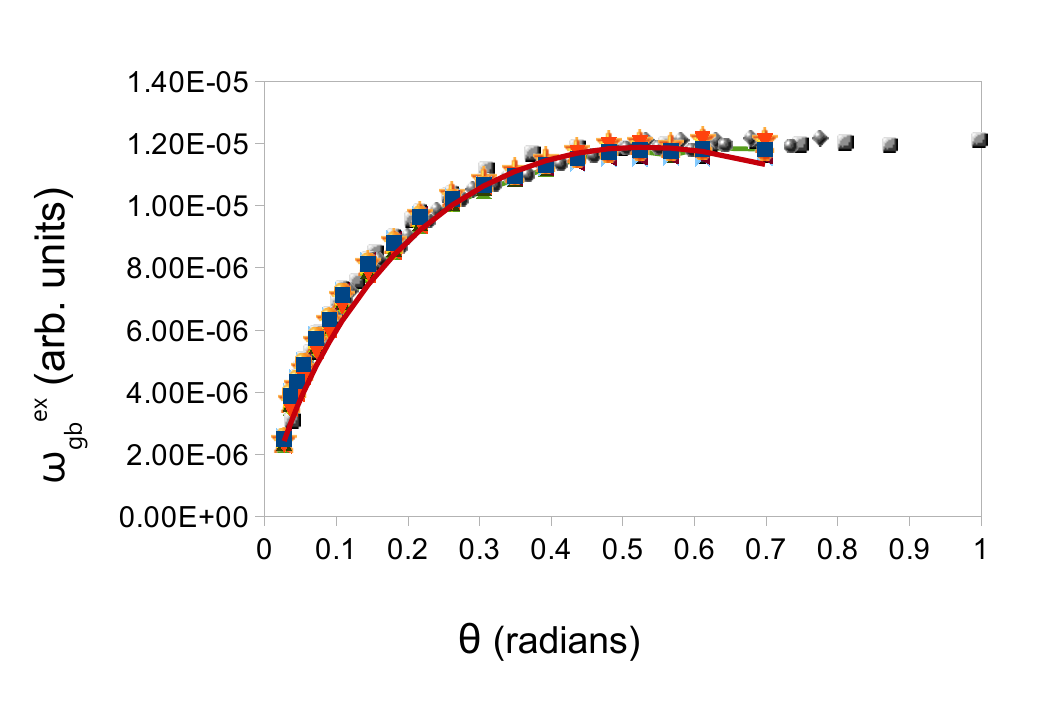}
	\caption{Scaled grain boundary energy for the PFC model. Plotted are $\omega_{gb}=\gamma_{gb} E^{m}_{0}(B^L_0=1.002,\psi_0=0)/(w_{gb} \Delta x E^{m}_{0}(B^L_0,\psi_0))$, vs. bicrystal crystal misorientation, $\theta/\theta_c$ (in radians for the reference curve) for different concentrations and temperatures.  Empirical reference Read-Shockley curve (solid). Data for $w=0.088$: $\psi_0=0$, $B^L_0=1.002$ (squares); $\psi_0=0$, $B^L_0=0.962$ (inverted triangle); $\psi_0=-0.05$, $B^L_0=0.996$ (right triangle); $\psi_0=-0.05$, $B^L_0=1.006$ (circle); $\psi_0=0.1$, $B^L_0=1.015$ (left triangle); $\psi_0=0.1$, $B^L_0=0.995$ (star); $\psi_0=-0.15$, $B^L_0=1.035$ (bow tie); $\psi_0=-0.15$, $B^L_0=1.015$ (triangle); $\psi_0=-0.15$, $B^L_0=0.995$ (diamond),$\psi_0=-0.2$, $B^L_0=1.045$ (graduated-shading circles). Data for $w=0.008$:  $\psi_0=-0.2$, $B^L_0=1.065$ (graduated-shading boxes), $\psi_0=-0.2$, $B^L_0=1.045$ (graduated-shading diamonds),$\psi_0=-0.2$, $B^L_0=1.025$ (line with ties).}
	\label{fig:gbenergy}
\end{figure}

We illustrate that grain boundary energy has a universal behaviour by scaling all raw data of grain boundary energy versus misorientation.     Furthermore, the shape of these curves can be approximated by a reference curve given by Eq.~\ref{eqReadShockley}, which is illustrated alongside the data in Fig.~\ref{fig:gbenergy}, where $E_0=1$ and $A=0.362$.  To match the simulated grain boundary energies to the reference curve, a vertical scaling factor $E'=E_0^m(B^L_0=1.002,\psi_0=0)(\Delta x)/E_0^m(B^L_0,\psi_0)/w_{gb} $ is applied, where $E_0^m$ are found empirically by fitting the data to Eq.~\ref{eqReadShockley}.   $E^m_0$ is related to the theoretical $E_0$ by a linear relation.  As in Fig.~\ref{fig:gbenergysmall}, the data points are scaled horizontally according to a stretch factor, $\theta_c=\exp(0.362-A)$.  For $w=0.088$ data, $A=0.362$ for all concentrations and temperatures studied. For $w=0.008$ data in Figure~\ref{fig:gbenergy}, $A$ ($0.719, 0.467, 0.411$) for  ($B^L_0=1.065,1.045,1.025$), respectively.  In all cases shown in Fig.~\ref{fig:gbenergy}, $\eta=0$. 


\begin{figure}
	\includegraphics[scale=0.3]{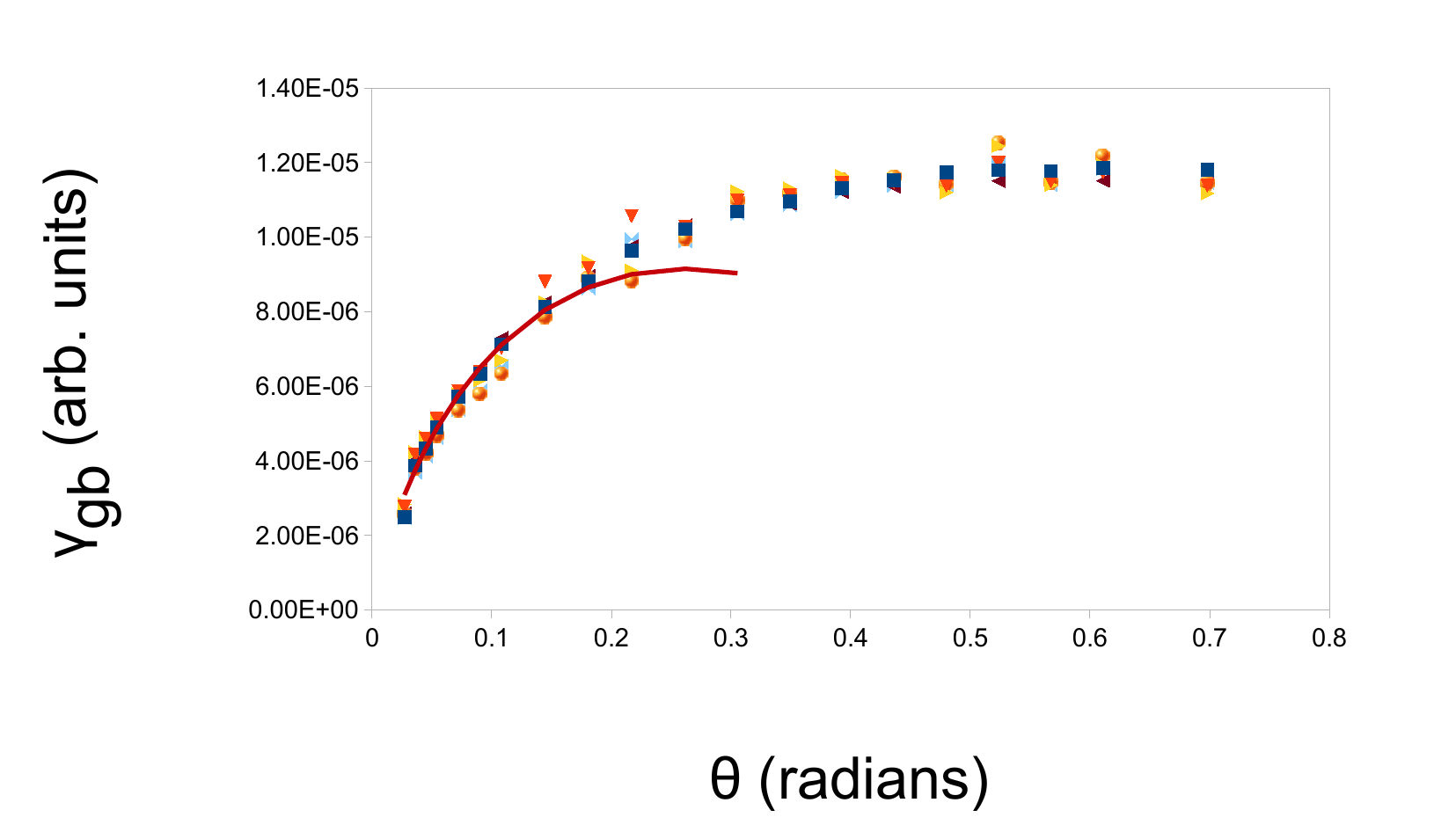}
	\caption{Grain boundary energy normalized by the system elastic constants for different degrees of mismatch and bonding energies in the PFC model.  Blue squares, maroon triangles - $\eta=0$; cyan ties, red triangles - $\eta=0.05$, yellow triangles, orange shaded balls - $\eta=0.1$.  We include a a fit to the low angle Read Shockley for comparison.}
		\label{fig:gb_energy_eta}
\end{figure}

To study the effect of the degree of mismatch,  $\eta$, on the thermodynamics of grain boundaries, we analyse runs at average concentration $\psi_0=-0.15$ for $w=0.088$.  In Fig.~\ref{fig:gb_energy_eta}, we plot  grain boundary energies normalized to the elastic constants of the system against the elastic constants at $B^L_0=1.002, \psi_0=0$.  Grain boundary energy is virtually indistinguishable for the non-zero $\eta$ values studied, except for some misorientations where the mismatch strains the structure a fair bit locally (most noticeable for $B^L_0=1.035,\eta=0.05$ at $4.13-8.28^\circ$, but also noticeable for $\eta=0.1$ at some of the higher angles), thereby increasing the free energy of the system.  For $8.28^\circ$ it was observed that for larger $\eta$ that the boundary can buckle, thereby relieving some of the stress. The $\eta \ne 0$ case is observed for this model to give only a small change in total interface segregation compared to the $\eta=0$ cases. This occurs because $\eta$ does not change the amplitude of the density field much, meaning that the Read-Shockley prefactor should be approximately the same.  This, in turn, lead to similar grain boundary energies for the two cases, as discussed further in Section \ref{secUndercooling}.

\subsubsection{Characterizing Solute Segregation in the PFC  Model}
\label{secPFCOrigSegregation}

To chracterize solute segregation, Eqs~(\ref{eqnevolA},\ref{eqpsievolA}) were directly simulated.   Because it was found that using a 1024x1024 grid gives comparable results to a 2048x1024 grid, these simulations were done with a grain boundary length of 1024 instead of 2048.  Simulations were done with $B^L_0=1.002$, $w=0.088$, $\mu_n=-0.25$  (meaning $n_o \neq 0$), and $\eta=0$, while $\mu_{\psi}$ was allowed to vary.  All other parameters were the same as in the previous section, namely, $B^L_2=-1.8$, $B^X=1$, $t=0.6$, $v=1$, $u=4$, and $K=4$.

\begin{figure}
	\includegraphics[scale=0.4]{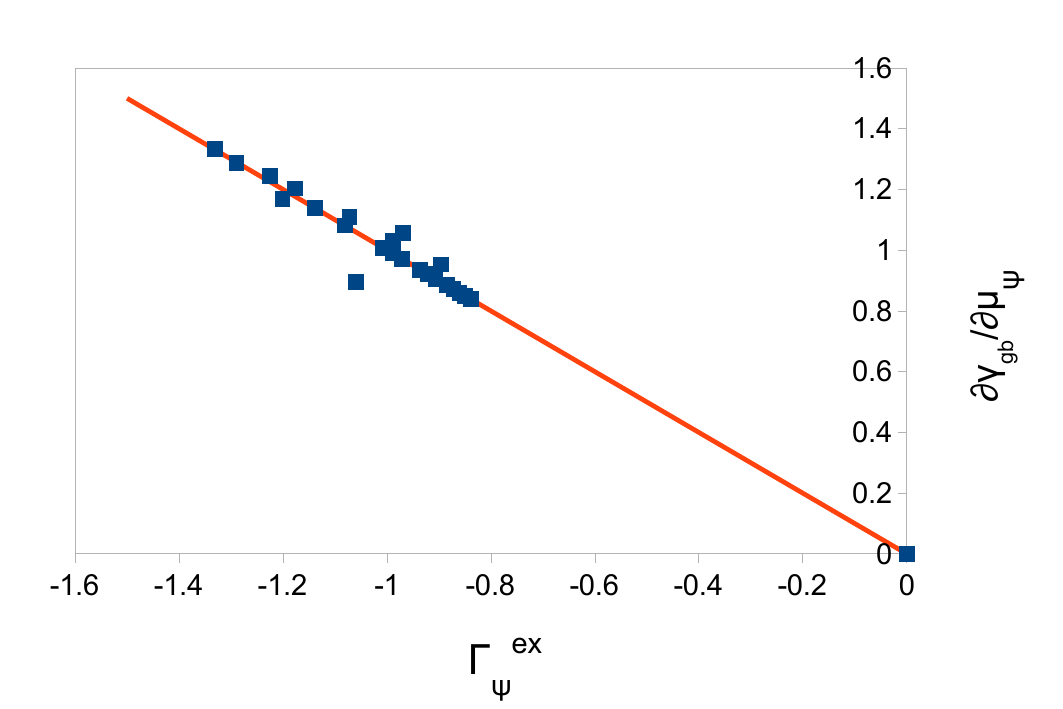}
	\caption{$\left(\partial \gamma_{gb} /\partial \mu_\psi\right)_{B^L_0,\mu_n}$ compared against simulated grain boundary excess concentration, $\Gamma^{ex}_\psi$.  A reference line with a slope of -1 is included.}
		\label{fig:Gibbsadsdirect}
\end{figure}

We determine the grain boundary energy with a method similar to \cite{mellenthinphasefield2008}, but only use two system widths perpendicular to the grain boundary, namely $N_x=1024$ and $N_x=2048$, to determine the grain boundary energy via eq~(\ref{eqgrandpot}).  We perform the simulation changing $\mu_{\psi}$ from $-0.18$ to $0$ in increments of 0.05.  We compute both sides of eq~(\ref{eqGibbsads2}), with $x=\psi$, numerically by
\[
\frac{\gamma^{i+1}_{gb}-{\gamma^{i-1}_{gb}}}{\mu_{i+1}-\mu_{i-1}}=-\Gamma^{ex,i}_\psi
\]
where the index $i$ represents these quantities for a certain chemical potential increment, $i$.  Results comparing the numerically computed $\left(\partial \gamma_{gb} /\partial \mu_\psi\right)_{B^L_0,\mu_n}$ with direct calculation of $\Gamma^{ex}_\psi$ are shown in Fig.~\ref{fig:Gibbsadsdirect}.  An analytic form for determining the total solute segregation in the PFC model is shown in the Appendix.

\begin{figure}
	\includegraphics[scale=0.4]{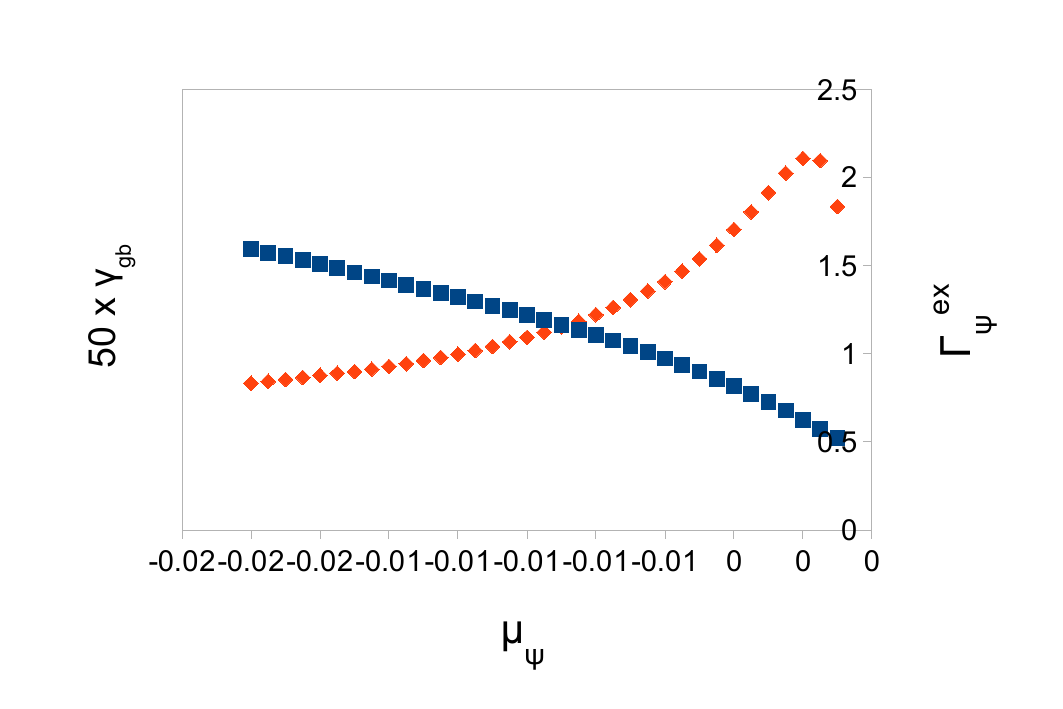}
	\caption{Computed grain boundary energy, $\gamma_{gb} $, (blue, squares) and excess grain boundary concentration, $\Gamma^{ex}_\psi$, (red diamonds)  vs chemical potential $\mu_\psi$ for the PFC model.}
		\label{fig:Gibbsadsconcener}
\end{figure}

Often in experimental papers of segregation, such as the work by Hondros and Seah \cite{cahnphysical1983}, segregation relation is demonstrated by superimposing how grain boundary energy and excess solute vary against chemical potential on the same plot.  This is shown in Fig.~\ref{fig:Gibbsadsconcener}.  A large decrease is seen in the free energy, due mostly to the change in elastic constants of the system.  This result will be examined further in the discussion.  Although the PFC model gives physically consistent  results, it does not describe dilute systems well, which is where this phenomenon is mostly studied experimentally.   

\subsection{XPFC Model}
\label{secNumXPFC}
\subsubsection{Method}
\label{secXPFCMethod}
To analyse our simulations, we use equations Eq. (\ref{eqnevolX}) and Eq. (\ref{eqcevolX}) to analyse how grain boundary energy varies with misorientation. These equations are simulated using the following semi-implicit method:

\beqa
\hat{n}^{t+1}_k=\frac{1}{1+p k^2\Delta t}(\hat{n}^{t}_k-\Delta t k^2 (\hat{\mu_n}^{t}_k-L_k p\hat{n}^{t}_k)) \nonumber \\
\hat{c}^{t+1}_k=\frac{1}{1+ \alpha k^4\Delta t}(\hat{n}^{t}_k-\Delta t k^2 (\hat{\mu_c}^{t}_k- k^2\alpha \hat{c}^{t}_k)) \nonumber
\eeqa
where $p=2$.  For determining how grain boundary energy changes with chemical potential, the following semi-implicit scheme is used to solve Eq. (\ref{eqnevolXA}) and Eq. (\ref{eqcevolXA}):

\beqa
\hat{n}^{t+1}_k=\frac{1}{1+\Delta t}(\hat{n}^{t}_k-\Delta t  (\hat{\mu_n}^{t}_k-\hat{n}^{t}_k)) \nonumber \\
\hat{c}^{t+1}_k=\frac{1}{1+ \alpha k^2\Delta t}(\hat{n}^{t}_k\Delta t k^2 (\hat{\mu_c}^{t}_k- k^2\alpha \hat{c}^{t}_k)) \nonumber
\eeqa

We use $\alpha=1$, $\eta=1.4$, $\chi=1$, $\omega=0.02$, and $c_o=0.5$.    For the correlation function, both species have: $G_0=8/81$, $D_0=1$, $P_0=-2$, $D_1=1/12$, $G_1=25/32$, and $P_1=1$.   The reciprocal of the lattice spacing is $k_1=2\pi$ for a system consisting entirely of material 1 ($c=0$) and  $k_2$ for $c=1$.   $k_2$, $\mu_n$, $\mu_c$, $\lambda$, $\Delta t$, $\sigma$, and $M_c$ vary depending on the simulation.  With the appropriate choices of $k_2$ and $\lambda$, we can simulate a eutectic phase diagram with varying solute solubilities.  We perform these simulations on a 1024x1024 grid.  Grain boundaries are formed in the same way that they were with the previous PFC model.  The lattice spacing is usually taken to be $dx = 0.125$ because the composition is close to $c=0$ (so that there are roughly 8 grid points between atomic planes).  A similar criterion to that which was used for the PFC model was used to determine equilibration for the XPFC model; for systems with large mismatch, the value of $s_{\mu n}$ needs to roughly $10^{-6}$ to ensure one is in the equilibrium.  The concentration field equilibriates much faster than the density field.

\subsubsection{Change of Grain boundary energy with misorientation}
\label{secXPFCGBEnergy}
We simulate the grain boundary energy in a system at an average concentration of $c=0.015$ and average density of $n=0$.  To make the lattice parameter of the solute atoms roughly $10\%$  larger than the solvent atoms, $k_2=9/5\pi$ was chosen.  A time step of $\Delta t = 1.0$ was used.  The same angles were chosen as in section \ref{secPFCOrigGBenergy}. The simulations were initially quenched to $\sigma=0$ and the temperature was raised and held at $\sigma=0.025,0.05,0.075,0.10,0.125$ for 100000 time steps.

\begin{figure}
	\includegraphics[scale=0.4]{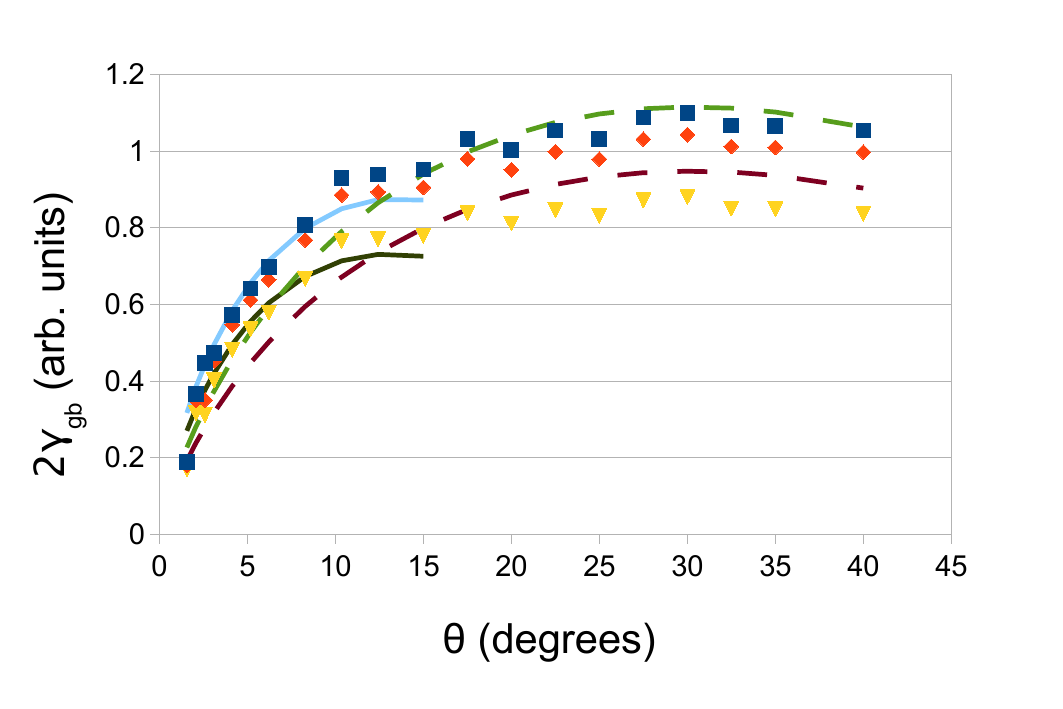}
	\caption{Grain boundary energy vs. misorientation for the XPFC model. Shown here are average system concentration $c=0.015$ for $\sigma=0.0$ (blue squares), $\sigma=0.05$ (red diamonds) and $\sigma=0.1$ (yellow triangles).  The two solid curves are shown to illustrate  the low-angle Read-Shockley fit to the $\sigma=0$ and $\sigma=0.1$ data. The dashed curves are included to illustrate that the high-angle Read-Shockley fit describes to the $\sigma=0$ and $\sigma=0.1$ data.}
		\label{fig:XPFCRS}
\end{figure}

For this set of simulations we use Eq.~(\ref{eqgrandpot1}) to determine the grain boundary energy.  Unlike the case of the PFC model, it is not as straightforward here to fit the Read-Shockley of  Eq.~(\ref{eqPFCelast}) to the low angle data.  Deviations from the smooth curves of the PFC model, when the mismatch $\eta \neq 0$, are even more pronounced in the XPFC binary model.   Because of this feature of the data, we determine bulk energy from Eq.~(\ref{eqgrandpot}) by tracking the system free energy dependence with $L_x$.  Fig.~\ref{fig:XPFCRS} shows the change of grain boundary energy versus mis-orientation for different temperatures. Only $\sigma=0.0, 0.05, 0.1$ are shown for clarity.    For reference, we show low and high angle Read-Shockley fits to the data.   We note that for low angles, fits yield a much smaller $E_0$ than the theoretical value of $Y_2 b /(8 \pi \alpha \phi^2 w_{gb})$.  This s because in the XPFC model, solute segregation decreases grain boundary energy significantly due to atomic mismatch.

\subsubsection{Characterizing the effect of solute segregation}
\label{secXPFCSegregation}

In this section, we characterize the XPFC model in same way as the PFC model in section \ref{secPFCOrigSegregation}, except that $c$ replaces $\psi$. We use non-conserved dynamics  Eqs. (\ref{eqnevolXA}, \ref{eqcevolXA}). Two sets of runs were done. For all runs with $\lambda=0$ and $\sigma=0$, we use $\Delta t=0.5$, $\mu_n=-0.2$ and $M_c=1.0$, while $\mu_c$ varied between $-0.024$ and $0.051$ in increments of $\pm 0.01$ and were equilibrated at each $\mu_c$ for $20000$ timesteps, following an initial condition that was equilibriated for 200000 timesteps at $\mu_c=0.016$.  For $\lambda=0.2$, $\sigma=0$, $\Delta t=0.25$, $M_c=1.0$, $\mu_n=-0.2$ and $\mu_c$ was varied between $-0.0225$ and $0.050$ in increments of $\pm 0.002$, equilibrated at each $\mu_c$ for $20000$ timesteps, after an initial condition that was relaxed for 400000 timesteps at $\mu_c=0.0$.

\begin{figure}
	\includegraphics[scale=0.4]{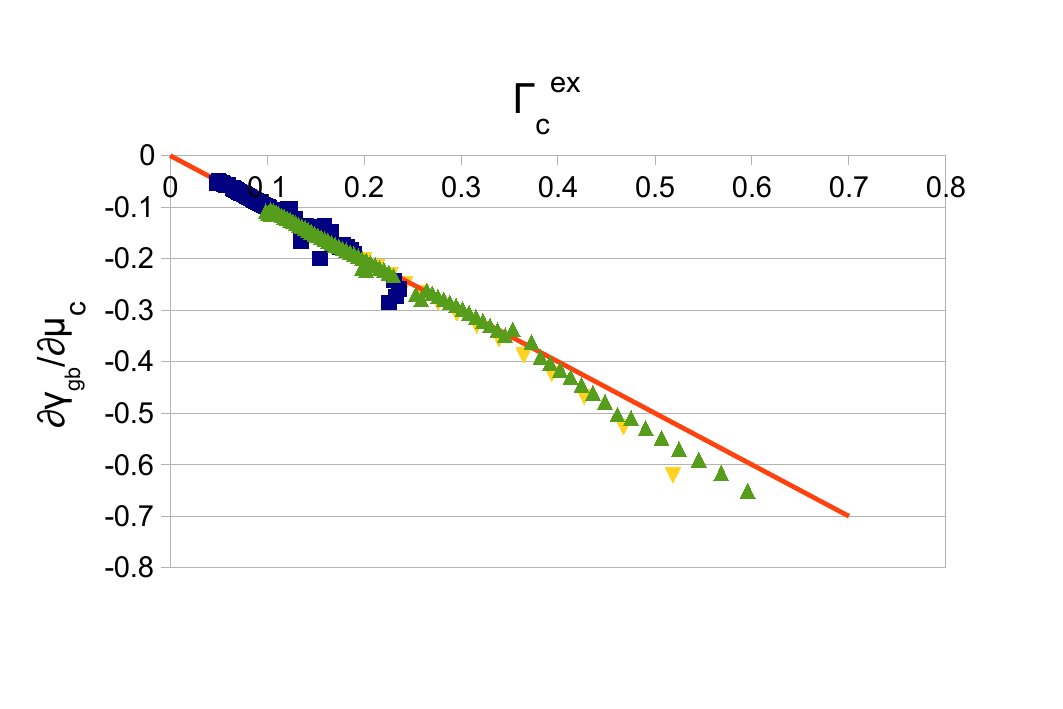}
	\caption{Prediction of $\Gamma_{c}^{ex}$ from Eq.~\ref{eqGibbsadsXPFC} vs. the same quantity obtained by direct numerical simulation of the XPFC model.  Green triangles - $\lambda=0$, misorientation angle $27.8^\circ$, Yellow triangles - $\lambda=0.2$,  misorientation angle $27.8^\circ$, Blue squares  - $\lambda=0$, misorientation angle $4.14^\circ$ . A reference line with a slope of -1 is included.}
		\label{fig:XPFCgibbsads}
\end{figure}

For this model, we analysed the following form of Gibbs adsorption theorem (cf. Eq~(\ref{eqGibbsads2})):
\begin{equation}
	\left(\frac{\partial \gamma_{gb}}{\partial \mu_c}\right)_{\sigma,\mu_n}=-\Gamma^{ex}_{c}
	\label{eqGibbsadsXPFC}
\end{equation} 
A comparison using the approach of Section~\ref{secPFCOrigSegregation} is shown in Fig.~\ref{fig:XPFCgibbsads}.   The deviation of the slope from $-1$ is likely due to numerical error in the estimation of the derivative.  This found that this error can be minimized by using a 5 point stencil to estimate derivatives, and by allowing longer equilibration times.  Some data were omitted in Fig.~\ref{fig:XPFCgibbsads} because the large mismatch of species 1 and 2 lead to  very large strains in the system, which caused the grain boundary to move in certain instances, leading to very long equilibration times. 

\begin{figure}
	\includegraphics[scale=0.4]{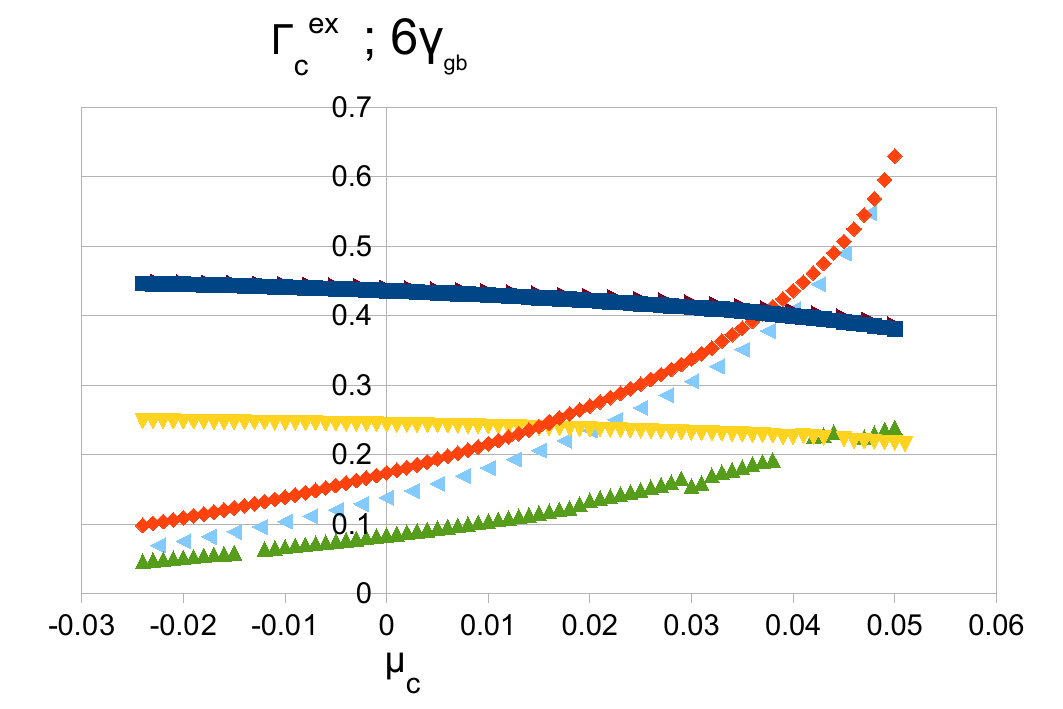}
	\caption{$\Gamma_{\psi}^{ex}$ and $\gamma_{gb}$ vs. $\mu_c$ for  the XPFC model. Shown for $\gamma_{gb}$ are: $\lambda=0$, misorientation angle $27.8^\circ$ (blue squares);  $\lambda=0$, misorientation angle $4.14^\circ$ (yellow upside-down triangles); $\lambda=0.2$, misorientation angle $27.8^\circ$ (maroon right-pointing triangles).  Shown for $\Gamma_{\psi}^{ex}$ are:  $\lambda=0$, misorientation angle $27.8^\circ$ (red diamonds);  $\lambda=0$, misorientation angle $4.14^\circ$ (green upward pointing triangles); $\lambda=0.2$, misorientation angle $27.8^\circ$ (cyan left-pointing triangles). }
		\label{fig:XPFCsegener}
\end{figure}

We also examined how grain boundary energy and excess concentration vary with $\mu_c$ for the XPFC model. This is shown in Fig.~\ref{fig:XPFCsegener} for the case of constant $\Delta x$ data.  We observe that a smaller degree of segregation is observed if the misorientation angle is lower.  This can be understood in terms of there being less amorphous, or disordered, material at the interface for the smaller misorientations.  It is noteworthy that  although the  $\lambda=0.2$ data is at a significantly lower {\emph average} concentration (roughly a factor of 3) than the $\lambda =0$ data, their amount of segregation to the grain boundary is comparable at equal chemical potentials. As expected when there is an enthalpy of mixing term which encourages phase separation, solute is much more strongly attracted to the interface, as observed $\lambda=0.2$ case, which is indicated by equal amounts of solute segregation to the interface as the $\lambda=0$ case, despite there being less total solute available in the entire system.  This is also noted by the grain boundary energies being comparable and changing at roughly the same rate.  

The segregation behaviour and grain boundary energy in Fig.~\ref{fig:XPFCsegener} are qualitatively similar to the corresponding plot in \cite{cahnphysical1983} for low concentrations (that is, low chemical potentials).  The increase in solute segregation continues here even for the largest chemical potentials, when the solid phase crosses the solvus line of the phase diagram and becomes metaphase.  No corresponding phase boundary is encountered for the largest chemical potentials in the data of  \cite{cahnphysical1983}.  It is noted that discontinuities in the solute segregation curve in Fig.~\ref{fig:XPFCsegener} for the $4.14^\circ$ simulations relate to the instances which were omitted from Fig. \ref{fig:XPFCgibbsads} because the system had not quite equilibriated yet.

\section{Discussion of Results}
\label{secDiscussion}

\subsection{Effect of Parameters on Grain Boundary Segregation}
\label{secUndercooling}

\begin{figure}
	\includegraphics[scale=3]{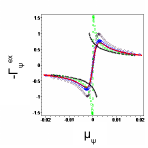}
	\caption{Segregation at misorientation of $22.5^o$ for different $\mu_\psi$ using Eq.~(\ref{eqGibbsadsPFC2}).  Red line- $B^L_0=0.01$; Blue circle- $\eta=0.1$,  $B^L_0=0.01$; Purple diamonds- $\eta=0.2$,  $B^L_0=0.01$; Green square-  $B^L_0=-0.01$; Black crosses-  $B^L_0=0.01$. In all cases $w=0.008$.}
	\label{fig:segregation_theory}
\end{figure}

Using Eq.~\ref{eqReadShockley} for the PFC model, we can derive an expression for the total amount of segregation based on Gibbs' adsorption theorem,  Eq. \ref{eqGibbsads}, and determine how segregation depends on material properties via the parameters of the PFC model.   We elucidated the effect of undercooling, chemical potential (concentration), mismatch, and energy of mixing on solute segregation to grain boundaries.  A typical plot $\Gamma_\psi^{ex}$ for various model parameters is shown in Fig.~\ref{fig:segregation_theory}. The plots were made using the analytic expressions in Eq.~(\ref{eqGibbsadsPFC2}), derived in the appendix and verified against simulations.   We note that for the case of $w=0.088$, the analytic curves resemble the segregation trends in Fig.~\ref{fig:Gibbsadsconcener}.   

Temperature ($\Delta B_0$) has a strong effect on the degree of segregation, as seen in the difference between the red and green curve.  This is due to $\Delta B_0$ strongly affecting the density amplitude, $\phi$, which in turn also affects $\mu_\psi$.  This trend is also found in the XPFC model (Fig.~\ref{fig:XPFCRS}), where temperature $\sigma$ has a strong effect on the elastic constants, which in turn determines the grain boundary energy, which in a similar vein affects the degree of segregation.  Far away from the phase boundary, the effect of undercooling (at constant $\sigma$) on segregation is very small.  However, close to the phase boundary changing undercooling by small amounts has a drastic effect on segregation, as shown at constant  $\sigma$ in Fig.~\ref{fig:XPFCsegener}.  

Figure~\ref{fig:segregation_theory} shows that lattice mismatch has a relatively small effect in the PFC model for small $|\eta|$ ($<0.1$), as seen by the small difference between blue and red curves.  However, the effect is larger for $\eta=0.2$, as seen by comparing the purple and red curves. Physically, we could expect a large size mismatch between atoms, $\eta$ to alter grain boundary segregation because the strains induced in the system by mis-matching local lattice constant increases the elastic free energy. To reduce free energy, solute atoms typically segregate to the grain boundary, which is typically less-ordered than the crystal.  For smaller mismatch, the effect is not easily discernible.  The small effect for $\eta \le 0.1$ can be explained in terms of $\phi^2$ and $\partial \mu_\psi/\partial \psi$ not changing substantially with a small change in $\eta$ (these quantities depend on $\eta^2$). For the XPFC model, the effect of the one lattice mismatch value studied was larger than for the corresponding misfit value in the PFC model. 

We found that eutectic systems in the PFC model ($w=0.008$) exhibit stronger segregation near $\mu_\psi=0$ or at the same concentrations $\psi_s$ than those with a lens-shaped (or double-lens-shaped ($w=0.088$)) phase diagram. This is due to the parameter $w$ in the binary PFC model, which is related to the difference of interspecies bond energy $\epsilon_{12}$ and self bond energies $\epsilon_{11}$, $\epsilon_{22}$  ; that is $w\sim(2 \epsilon_{12}- \epsilon_{11}-\epsilon_{22} )$, where $1$ and $2$ represent the two kinds of atoms present.  As $w$ becomes smaller, the self-attraction becomes larger, meaning that atoms of type 1 want to be near type 1 atoms and type 2 atoms near type 2.  This implies a larger degree of segregation is more energetically favourable in such alloys.  This is consistent with what Hondros and Seah who related solute enrichment factors (at the grain boundary) with solute solubility (which is related to $2 \epsilon_{12}- \epsilon_{11}-\epsilon_{22}$) \cite{sutton_interfaces_1995, cahnphysical1983}.  We observe the same trend in the XPFC model by comparing the $\lambda=0$ and $\lambda=0.2$ cases, where the $\lambda=0.2$ case has lower solute solubility.  Because the degree of solute segregation at a given chemical potential is similar for both  cases,  but the average concentration is much lower in the  $\lambda=0.2$ case,  the segregation per solid solubility (that is, the enrichment factor) is larger for the $\lambda =0.2$ case. 

Another interesting property of the curves in Fig.~\ref{fig:segregation_theory} is that a number of the curves for the PFC binary model have both a maximum and a minimum degree of segregation (concentration excess).  We would expect curves in a normal material to have at least one extremum because the excess solute is zero in a pure material and non-zero value when there is a mixture;  this observation does not contradict Hondros and Seah, who predict a monotonic increase in surface excess Fe-P with increase P concentration because they only consider small concentrations, where Henry's Law is used to determine the chemical potential \cite{sutton_interfaces_1995}.  The XPFC binary model studied in this paper also shows monotonically changing grain boundary segregation with average system concentration (chemical potential) changing.  The XPFC model also obeys Henry's law at small concentrations.
  
The two extrema (one minimum and one maximum) in the PFC model is partially a result of the alloy behaving mostly like a pure material for $\psi_0=0$, so there is one extremum between the pseudo-pure (50$\%$ composition) and each of the pure materials (0$\%$ and 100$\%$ composition).  Nonetheless qualitatively similar behaviour would be expected for other completely miscible alloys.  That is, when the material is species 1 rich, there could an accumulation of species 2 at a grain boundary and an accumulation of species 1 at a grain boundary when the material is species 2 rich; unless the change is discontinuous, these two extrema must be observed.  For more complex materials, once we specify an appropriate formalism, Eq. \ref{eqGibbsads} can be used to at least numerically predict where this extremum will occur.   We also observe that the segregation peaks shift with temperature.   The grain boundary can be considered as an amorphous material, which becomes more metastable relative to the solid as the system is undercooled, thereby shifting the equilibrium concentration between these phases nearer $\psi_0=0$.

\subsection{Effect of Parameters on Grain Boundary Energy}
\label{secDiscGBEnergy}

There are two main considerations to make when studying how grain boundary energy is affected by solute segregation: the extent by which the alloy grain boundary energy differs from a pure material and the extent to which grain boundary energy changes in a given system as thermodynamic variables are changed---temperature or chemical potential, for example.  For the PFC model with $w=0.088$, we observed that the the grain boundary energy in the alloy does not differ much from that of a pure material with equal elastic moduli.   For low angles, we might expect this behaviour from  Turnbull's estimation of the interface energy for systems with low lattice mismatch (that is, a boundary consisting of discrete dislocations),  in which the elastic and not the chemical component of the grain boundary energy is the dominant effect \cite{howe_interfaces_1997}.   Although the core size changes slightly as the undercooling changes, as shown in Fig.~\ref{fig:gbcoreradius}), because the cores make up only a small fraction of the volume of the grain boundary, the grain boundary energy for an alloy does not significantly differ from that of a pure material, which we calculate from the datasets for $\psi_0=0$ ---which are mathematically equivalent to simulating a pure material. 

On the other hand, the grain boundary energy for high angles is expected to be roughly constant \cite{sutton_interfaces_1995} and dominated by the presence of an undercooled metastable phase between the crystals \cite{Warren20036035}.  This metastable phase is present throughout the entire grain boundary and so solute segregation might be expected to show a  stronger effect for high angle grain boundaries than for low angle grain boundaries.  Eq.~\ref{eqGibbsadsPFC2} confirms that the degree of solute adsorption, $\Gamma^{ex}_{\psi}$, increases with angle because the larger angles have higher grain boundary energies.  Yet for the PFC model with $w=0.088$, the change in grain boundary energy relative to the pure material at the same undercooling is small as indicated in Fig.~\ref{fig:gbenergy}.  That being said, the grain boundary energy can still undergo large changes with chemical potential changing with temperature, as indicated in Figs.~\ref{fig:Gibbsadsconcener} and ~\ref{fig:segregation_theory} .

For the PFC parameter $w=0.008$, which simulates eutectic phase diagram in the PFC model, grain boundary energy changes significantly when the bonding between atoms changes.  For eutectic simulations near the spinodal line, solutal effects noticeably depress the grain boundary energy, $\gamma_{gb}$.  This result is further supported by simulations in the XPFC model when the two types of atoms are not very soluble in each other.  In Fig.~\ref{fig:XPFCRS},  the Read-Shockley curve is fit with a much smaller than that expected for the pure material.  Although the grain boundary energy does not decrease as much for the XPFC as for the PFC model when chemical potential changes (since elastic constants are roughly constant at constant $\sigma$ for the average concentrations studied), meaning that the decrease in $\gamma_{gb}$ is almost entirely due to the solute segregation.

Comparing our results with some experimental and molecular dynamics studies,  we see that the effect of solute segregation on grain boundary energy can be quite large when the atomic species are rather immiscible \cite{sutton_interfaces_1995, Kirchner2011406, Shibuta20091025} or quite small when they are miscible (e.g. Fe-Cr at high temperature \cite{Shibuta20091025} or the double lens-shaped phase diagrams, e.g. Cu-Au \cite{cahnphysical1983}).   When this effect is typically measured, the particular mechanism for a decrease in grain boundary energy is not necessarily known.  An advantage of PFC models over experiments is that the elastic constants and other properties of the system can be easily determined and we can isolate what mechanism is responsible for a change in grain boundary energy.  For example studying the PFC model with $w=0.088$, we learn that the strong decrease in grain boundary energy with changing chemical potential near $\psi_0=0$, as shown in Fig.~\ref{fig:Gibbsadsconcener}, is predominantly related to the elastic constants changing, as opposed to say the structure changing significantly.  That is to say, this change in grain boundary energy is mainly due to undercooling (which, of course, is a function of $\psi$ at constant $B^L_0$), {\it  rather than} solute segregation directly for these model alloys. 

We make one last note about the Read-Shockley form being roughly followed by all grain boundary energy curves that were studied. This finding is not surprising given that the base contribution of $\gamma_{gb}$ has the Read-Shockley form and that the decrease in energy due to solute segregation is proportional to $\Gamma^{ex}_{x}$, which according to Eq.~\ref{eqGibbsads} is proportional to $\gamma_{gb}$.  Because the various parts determining $\gamma_{gb}$ have a Read-Shockley form,  $\gamma_{gb}$ should maintain this form despite the stronger solutal effects in the eutectic PFC model and XPFC model.

\subsection{Future Extensions of PFC model}
\label{secExtension}

It should be noted that the PFC model contains only the essential effects of elasto-plasticity and thermodynamics on atomistic length scales. This work has shown that the PFC alloy models is nevertheless able to capture the basic details of grain boundary segregation and energy and their dependence on undercooling, average alloy concentration, mismatch, and energy of mixing.  With new improvements that take into account correlations between atoms more accurately, more realistic PFC models being developed \cite{wu_controlling_2010, PhysRevE.81.061601, greenwood_free_2010, Greenwood2011, PhysRevE.80.031602, Berrydefect2,Ofori10} can display more accurate  segregation properties, as was shown here for the XPFC model. Further improvement of the PFC and XPFC formalism have similarly been developed that further capture other properties. For example, the work of Berry et al demonstrates that the properties of defects can be more realistically simulated if higher order terms are included in the correlation function \cite{Berrydefect2}.

A key motivation for this study was to quantify how grain boundary energy and segregation could be quantitatively controlled in two PFC models used in the literature presently. Solute segregation to grain boundaries can have a significant effect on dendritic growth, because even small changes in surface energy anisotropy yield measurable changes in microstructure morpholog.  Another property that is expected to be more strongly affected by segregation to grain boundaries and around dislocation cores is grain boundary pre-melting behaviour, in which a system might display multiple grain boundary widths at the same state variables. In fact, Gibbs adsorption theorem, as stated in \cite{sutton_interfaces_1995}, should even be applicable for wetted grain boundaries, which can be a powerful tool for studying premelting in alloys.

\section{Conclusion}
\label{secConclusion}
Both PFC alloy models studied here have been shown in numerous works to self-consistently capture the thermodynamics and elasto-plasticity inherent in many diffusive phase transformations in metals. This work provided another test of the robustness of the PFC formalism in predicting the important physical properties of solute segregation and grain boundary energy in binary alloys.

We used two phase field crystal alloy models to study solutal effects on grain boundary properties in spinodal and eutectic binary alloys.  We derived a semi-empirical model of excess solute segregation to the grain boundary. The role of undercooling, average alloy concentration, lattice misfit on the grain boundary energy on spinodal and eutectic alloys was characterized for the original PFC alloy model, and compared to the corresponding results computed from the XPFC alloy model, and experiments. We found that alloys with lens-shaped phase diagrams exhibit a negligible segregation effect on grain boundary energy, both at low and high mis-orientation angles. However, undercooling (or more generally temperature) strongly impacts the energy in PFC models, through their effect on the elastic coefficients.  This finding is in agreement with experiments, which find that binary materials having high solubility of one material in the other show little change in grain boundary energy with composition changes at dilute compositions --except at higher compositions where grain boundary energy changes more significantly, as was observed in the PFC model. For eutectic alloy systems, on the other hand, solute segregation to the grain boundary had a stronger direct impact on its energy, again consistent with other works.  The direct effect of solute segregation on grain boundary energy was particularly strong for the XPFC model, though the grain boundary energy did not vary much as the system average concentration changed at constant temperature.  We also found that (small) lattice mismatches (i.e. Vegard's law parameter $\eta$) did not strongly affect segregation, though we observe that higher degrees of mismatch can have a profound effect on segregation (which is consistent with the prediction of Eq.~(\ref{eqGibbsadsPFC2})) and the grain boundary energy, as is particularly clear in the XPFC model. 

There are numerous applications of the results and methodology found in this article.  The phase field crystal formalism can be linked to traditional mesoscale phase field methods through various coarse graining procedures 
\cite{PhysRevE.81.011602,PhysRevB.81.214201,PhysRevE.82.021605}.  
As a result, the grain boundary energy and segregation results inferred from this work can help guide the parameterization of mesoscale continuum theories whose forms are often---by necessity---phenomenological (e.g., \cite{Warren20036035,Ofori10}). Similar phase field phenomenologies have been recently used to make predictions about grain boundary wetting---in particular about how the disjoining pressure changes as the system parameters change  (e.g., \cite{Mishin20093771,PhysRevE.81.051601}).

\section{Acknowledgements}
The authors acknowledge the National Science and Engineering Research Council of Canada for funding.  We acknowledge Sharcnet (NcMaster) and Clumeq (McGill) High-Performance Computing Centres of Compute Canada for computational resources. We also thank Harith Humadi, Nana Ofori-Opoku, Jeff Hoyt, Ken Elder, Michael Greenwood, Joel Berry, and Peter Stefanovic for help with code development and useful discussions. 

\section{References}
\bibliographystyle{apsrev4-1}
\bibliography{PFC_references, bibliomaster_Nik}

\section{Appendix -- An alternate approach to Characterizing Solute Segregation in the PFC  Model}

In this section we present an analytic expression for how grain boundary energy and grain boundary segregation can be related to each other in the PFC model.  With a semi-analytical characterization for $\gamma_{gb}$ in hand, we can proceed with the derivation using the Gibbs' Adsorption 
Theorem defined by Eq.~\ref{eqGibbsads}.  We begin by noting that in terms of the notation used in the PFC alloy model, the expression for excess solute in 
Eq.~\ref{eqGibbsads} can be recast as
\beqa
	\left(\frac{\partial \gamma_{gb}}{\partial \mu_{\psi}}\right)_{T,p} &=& \left(\frac{\partial \gamma_{gb}}{\partial \mu_{\psi}}\right)_{T,V} 
	\!-\! V_{ex}\left(\frac{\partial p}{\partial \mu_\psi}\right)_{T,V} \nonumber \\ 	&=& 
	\left(\frac{\partial \gamma_{gb}}{\partial \psi}\frac{\partial \psi}{\partial \mu_{\psi}}\right)_{T,V} \!-\! w_{gb}\frac{n_s-n_g}{1+n_s}\psi 
	\nonumber \\ &=&
	-\Gamma^{ex}_{\psi}
	\label{eqGibbsadsPFC}
\eeqa
where it is understood that the formal thermodynamic variable $T$ is to be identified by and substituted by the reduced temperature variable $B_0^L$ of the PFC model, and $w_{gb}$ is the width of grain boundary region.  Both derivatives in Eq.~\ref{eqGibbsadsPFC} can be determined by making use of Eq.\ref{eqReadShockley} and  Eq.\ref{eqPFCelast}, and by noting that   $\mu_{\psi}=\partial f/ \partial \psi$,  where $f(\psi,\phi(\psi),B_0^L)$ is the free energy density derived from the PFC model from the single mode approximation of the density $n$. This has been derived in Ref.~\cite{PhysRevE.81.011602}, from which it can also be shown that $\mu_{\psi}=(w+6 B^L_2 \phi^2-24 \eta B^X_0 \phi^2 )\psi + u\psi^3$.  The theoretical quantity $\mu_\psi$ differs from the simulated $\mu_\psi$, which also implies that there is a systematic uncertainty on any calculated quantities which use this value.  Using the definition of $\mu_\psi$, we obtain,
\beqa
	&&\frac{\partial \psi}{\partial \mu_{\psi} }= \label{eqdpsidmu} \\ &&\frac{1}{ (w+(6 B^L_2 \phi-24 \eta B^X_0 \phi)(\phi+2\psi))+ 3 u \psi^2+ \partial \phi/\partial \psi} 
	 \nonumber
\eeqa
and
\beqa
	\frac{ \partial \gamma_{gb} }{\partial \psi} &=& \frac{\partial E^m_0}{\partial \phi} \, \frac{\partial \phi}{\partial \psi} 
	\, \theta \, \left(A-\ln(\theta)\right) 	\label{eqdphidpsi} \\ 	
	&=& 2\phi \frac{-0.80(B^L_2 -4 \eta^2)\psi \theta(A-\ln(\theta)) }{ \sqrt{t^2-15v(B^L_0-B^X+(B^L_2-4\eta^2)\psi^2)} } \nonumber
\eeqa
Substituting Eqs.~\ref{eqdpsidmu} and ~\ref{eqdphidpsi} into Eq.~\ref{eqGibbsadsPFC} yields 

\beqa
	&\Gamma&^{ex}_{\psi} =-w_{gb}\frac{n_s-n_g}{1+n_s}\psi \nonumber\\ &-&\frac{1.60 \, \phi}{(w+6 (B^L_2-4 \eta B^X_0) \phi(\phi+2\psi\partial \phi/\partial \psi ))+ 3 u \psi^2} \nonumber \\
	 &\times& \frac{((B^L_2-4\eta^2) \psi)\theta (A-\ln(\theta))}{\sqrt{t^2-15v(B^L_0-B^X+(B^L_2-4\eta^2)\psi^2)}} 	
	\label{eqGibbsadsPFC2}
\eeqa

The theoretical expression in Eq.~\ref{eqGibbsadsPFC2} is compared directly with numerical simulations of grain boundary segregation.  We follow Cahn's method \cite{sutton_interfaces_1995} for determining excess solute, using the  
variable $n$ and $\psi$. This yields an expression for the excess concentration in a grain boundary:
\beq
	\Gamma^{ex}_{\psi}=w_{gb} \left( \psi_{gb} - \frac{(1+n)_{gb}}{(1+n)_{s}}\psi_{s} \right)
	\label{eqExcess}
\eeq

\end{document}